\documentclass[11pt]{article}[12pt]
\usepackage[utf8]{inputenc}
\usepackage[english]{babel}
\usepackage{amsmath}
\usepackage{amssymb}
\usepackage{amsthm}
\usepackage{graphicx}
\usepackage{epstopdf}
\usepackage[affil-it]{authblk}

\usepackage{hyperref}
\hypersetup{
    colorlinks=true,
    linkcolor=red,
    citecolor=blue,
    urlcolor=blue
    }

\usepackage[a4paper, total={6in, 9.6in}]{geometry}

\begin{document}
\markboth{Petruhanov, V.N.; Pechen, A.N.}{Optimal control for state preparation in two-qubit open quantum systems}

\title{\LARGE\textbf{Optimal control for state preparation in two-qubit open quantum systems driven by coherent and incoherent controls via GRAPE approach}
}

\date{}

\author[1,2,*]{Vadim N. Petruhanov}
\author[1,2,**]{Alexander N. Pechen}

\affil[1]{\it \normalsize \href{http://mi-ras.ru/eng/dep51
}{Department of Mathematical Methods for Quantum Technologies},\par
Steklov Mathematical Institute of Russian Academy of Sciences,\par
8~Gubkina str., Moscow, 119991, Russia, }
\affil[2]{\it National University of Science and Technology ``MISiS'',\par
6~Leninskiy prospekt, Moscow, 119991, Russia;}
\affil[*]{vadim.petrukhanov@gmail.com, \href{http://www.mathnet.ru/eng/person176798}{mathnet.ru/eng/person176798}}
\affil[**]{apechen@gmail.com, \href{http://www.mathnet.ru/eng/person17991}{mathnet.ru/eng/person17991}}

\maketitle

\medskip

\rightline{\it\large Dedicated to the 75th anniversary of}
\rightline{\it\large the birth of Igor Vasyl'evich Volovich}

\medskip
\medskip

\begin{abstract}
In this work, we consider a model of two qubits driven by coherent and incoherent time-dependent controls. The dynamics of the system is governed by a Gorini--Kossakowski--Sudarshan--Lindblad master equation, where coherent control enters into the Hamiltonian and incoherent control enters into both the Hamiltonian (via Lamb shift) and the dissipative superoperator. We consider two physically different classes of interaction with coherent control and study the optimal control problem of state preparation formulated as minimization of the Hilbert--Schmidt distance's square between the final density matrix and a given target density matrix at some fixed target time. Taking into account that incoherent control by its physical meaning is a non-negative function of time, we derive an analytical expression for the gradient of the objective and develop optimization approaches based on adaptation for this problem of GRadient Ascent Pulse Engineering (GRAPE). We study evolution of the von Neumann entropy, purity, and one-qubit reduced density matrices under optimized controls and observe a significantly different behavior of GRAPE optimization for the two classes of interaction with coherent control in the Hamiltonian.

\end{abstract}

\allowdisplaybreaks

\section{Introduction} 
\label{Introduction}

Quantum control which aims to manipulate individual quantum systems is an important tool 
necessary for development of modern quantum 
technologies~\cite{Glaser_Boscain_Calarco_et_al_2015,Koch_arXiv2022}. In real experimental applications, controlled quantum systems are typically open, that is, interacting with their environment. This interaction with the environment is often considered as an obstacle for controlling the system. However, the environment can also be used to actively control quantum systems via its temperature, pressure, or more generally, non-equilibrium spectral density. 

For open quantum systems, in the works~\cite{Pechen_Prokhorenko_Wu_Rabitz_2008,Wu_Pechen_Rabitz_Hsieh_Tsou_2008,Oza_Pechen_Dominy_Beltrani_Moore_Rabitz_2009}, a general approach based on gradient optimization over complex Stiefel manifolds for quantum control and quantum technologies was developed. General case of arbitrary $N$-level quantum systems was considered, for which not only general framework of optimization over complex Stiefel manifolds was introduced, but also explicit analytical expressions for the gradient and Hessian of quantum control objectives for various quantum control problems were computed, the corresponding optimization techniques were developed, and various examples were studied, including with constraints.

The dynamical method of \textit{incoherent control} via  dissipation induced by an engineered environment with {\it time-dependent decoherence rates} was proposed in~\cite{Pechen_Rabitz_2006}. This method exploits density of particles of the environment $n_{\omega,\alpha}(t)$ in their momenta ${\bf k}$ (or energy which for incoherent photons is $\hbar\omega=\hbar c|{\bf k}|$, where $c$~is the speed of light and $\hbar$ is the Planck constant, or for massive particles of mass $m$ is $\omega=\frac{|{\bf k}|^2}{2m}$ ) and internal degrees of freedom $\alpha$, to induce time-dependent decoherence rates $\gamma_i(t)$ and drive the system density matrix towards a desired target state. Natural examples of such environment are the environment formed by incoherent photons and control through collisional decoherence; both approaches were studied in~\cite{Pechen_Rabitz_2006}. A particular case is control by temperature and pressure but generally the environment in this approach can be non-thermal and non-equilibrium and even time dependent. For incoherent photons, it is possible to realize various non-equilibrium
distribution functions $n_{\omega,\alpha}(t)$ (for photons $\omega$ is frequency and $\alpha$ is polarization). Incoherent control via decoherence and engineered dissipation used as a resource was studied in various contexts, e.g., for superabsorption of light via quantum engineering~\cite{HigginsNatComm2014}, optimal control for non-Markovian open quantum systems~\cite{HwangPRA2012} and Markovian dynamics~\cite{LucasRPL2013}, control of dissipation in cavity QED~\cite{LiningtonPRA2008}, manipulation of states of a degenerate three-level quantum system~\cite{Kozyrev2016}, incoherent control in a Bose-Hubbard dimer~\cite{ZhongPRA2011}, photoionization of atoms under noise~\cite{SinghPRA2007}, generating quantum coherence through an autonomous thermodynamic machine~\cite{MukhopadhyayPRA2018}, incoherent control of optical signals via quantum heat-engine approach~\cite{QutubuddinPRR2021}, optimization of up-conversion hues in phosphor~\cite{LaforgeJCP2018}, Landau-Zener transitions~\cite{Pechen_Trushechkin_2015}, etc. In this work, we study coherent and incoherent control of two-qubit quantum systems based on the type of master equations derived in the weak coupling limit by E.B.~Davies~\cite{Davies1976} and in the stochastic limit by L.~Accardi, Y.G.~Lu and I.V.~Volovich~\cite{Accardi2002}. Master equations beyond secular approximation are also considered~\cite{Trushechkin2021}.

In~\cite{Pechen_2011}, it was found that incoherent control by photons, when combined with coherent control by lasers, allows to approximately steer any initial density matrices, pure or mixed, of a generic $N$-level quantum system to a vicinity of any predefined pure or mixed target density matrix thereby to approximately, with some physical precision, realize complete density matrix controllability --- the strongest degree of quantum state control  --- of almost all (i.e., generic) quantum systems of arbitrary dimension. Important is that this result was obtained within physical class of master equations well known in quantum optics and that optimal incoherent control was found analytically. The control scheme is also independent of the initial state --- it allows to steer simultaneously all initial states into the same target state. Thereby this scheme realizes universally optimal Kraus maps~\cite{Wu_Pechen_Brif_Rabitz_2007}. 

Under coherent and incoherent controls in the Gorini--Kossakowsky--Sudarchhan--Lindblad (GKSL)
master equation, the exact degree of precision has been obtained recently only 
for a qubit~\cite{LokutsievskiyJPA2021}. Using geometric control theory, in this work it was found that most states in the Bloch ball can be obtained exactly except of points representing density matrices in two regions of the size $\delta\approx\gamma/\omega$, where $\gamma$ is the decoherence rate and $\omega$ is the transition frequency of the qubit. Moreover, the size of these regions was shown to be exactly in the range
\[
\frac{1}{2}\Bigl(1+\frac{\gamma^2}{\omega^2}\Bigr)^{-1/2}\le \delta\le\frac{\pi}{4}\frac{\gamma}{\omega}.
\]

Various numerical optimization schemes are used for quantum control including 
Pontryagin maximum principle~\cite{Boscain_Sigalotti_Sugny_2021},
steepest descent~\cite{Gross_Neuhauser_Rabitz_article_1992},
Krotov method~\cite{Tannor_Kazakov_Orlov_1992, Jager_Reich_Goerz_Koch_Hohenester_2014, Morzhin_Pechen_RussianMathSurveys_2019},
Zhu--Rabitz~\cite{Zhu_Rabitz_1998},
Maday--Turinici~\cite{Maday_Turinici_2003},
GRadient Ascent Pulse Engineering (GRAPE) method~\cite{Khaneja_Reiss_Kehlet_SchulteHerbruggen_Glaser_2005, 
SchulteHerbruggen_Sporl_Khaneja_Glaser_2011, Jager_Reich_Goerz_Koch_Hohenester_2014, 
Volkov_Morzhin_Pechen_JPA_2021},
genetic evolutionary algorithms~\cite{Judson_Rabitz_1992, Pechen_Rabitz_2006}, 
speed gradient~\cite{Ananevskii_Fradkov_2005, Pechen_RMS_2016}, 
Chopped Random-Basis (CRAB)~\cite{Caneva_Calarco_Montangero_2011}, 
Hessian based optimization~\cite{Dalgaard_Motzoi_et_al_2020}, etc. 
Uncomputability of discrete quantum control was shown via establishing a relation with Diophantine 
equations and tenth Hilbert problem~\cite{Bondar_Pechen_2020}. Numerical optimization schemes for a qubit driven by coherent and incoherent controls were 
studied in~\cite{Morzhin_Pechen_IJTP_2021_2019, 
Morzhin_Pechen_Physics_of_Particles_and_Nuclei_2020, Morzhin_Pechen_LJM_2020, 
Morzhin_Pechen_SteklovProceedings_2021, Morzhin_Pechen_AIP_Conf_Proc_2021, Morzhin_Pechen_arXiv2205.02521} 
for various objective criteria, for studying reachable and controllability sets, and for exploiting machine learning. 

While dynamics and coherent control of two qubits have been studied in a variety of works~\cite{Wang_Babikov_2011, Allen_Kosut_Joo_Leek_Ginossar_2017, Hu_Ke_Ji_2018, Bukov_Day_Weinberg_et_al_2018}, simultaneous coherent and incoherent control for the two-qubit case remains mostly uninvestigated. Global search genetic evolutionary algorithm was used for optimization of time-independent incoherent control in such systems driven by a GKSL master equation  in~\cite{Pechen_Rabitz_2006}.
The recent article~\cite{Morzhin_Pechen_LJM_2021} 
considers a two-qubit system with time-dependent coherent control and time-independent incoherent control. 

In this article, control of a two-qubit system is studied when both coherent and incoherent controls are modeled, in general, 
as {\it variable in time} piecewise continuous functions. For a fixed final time~$T$, we consider the problem 
of state preparation formulated as minimization of the Hilbert--Schmidt distance $\|\rho(T) - \rho_{\rm target} \|$  between the quantum system's final state, $\rho(T)$, which is found via solving the GKSL master equation,
and a given target density state $\rho_{\rm target}$. For convenience, we consider
the problem of minimizing the distance's square, i.e. $\|\rho(T) - \rho_{\rm target} \|^2$.
Under sufficient control resources 
(i.e., final time, possible additional constraints to controls' amplitudes, etc.), this problem describes steering a state of the system to the target state.

For this optimal control problem, we derive the following optimization technique
taking into account that incoherent control by its physical meaning {\it is always constrained} --- it is a non-negative function of time. Using piecewise constant controls, we introduce a change of variables and 
reduce the infinite-dimensional optimization problem to {\it unconstrained} 
finite-dimensional optimization, analytically compute gradient of the objective 
function and use it to adopt GRAPE, which is well known in NMR pulse sequence design, 
for numerical optimizing of coherent and incoherent controls. Here gradient 
of the objective function is derived analytically (for analogy 
see~\cite[subsections 5.2, 5.3]{Volkov_Morzhin_Pechen_JPA_2021}) via operations
with matrix exponentials. 

The structure of this paper is the following. In Sec.~\ref{Model}, 
the model of the two-qubit control system with the objective functional 
is formulated and standard definitions of other considered quantities 
such as von Neumann entropy, etc. are provided.
Sec.~\ref{GRAPE} describes the adaptation of GRAPE to the problem of
optimizing controls for minimizing the Hilbert--Schmidt distance's square, 
when piecewise constant controls are considered, and contains the corresponding 
numerical results for steering either mixed separable or Bell entangled state 
into a target mixed separable state. Conclusions section summarizes the paper.

\section{Control System, Objective Functional, and Relevant Quantities}
\label{Model}

Consider a pair of qubits (i.e., two-level quantum systems) interacting with the environment. 
Hilbert space of each qubit is ${\cal H}_i=\mathbb C^2$, $i=1,2$. Density matrix of the system 
$\rho$ is a $4\times 4$ positive semi-definite matrix with complex elements and with unit trace, 
i.e. $\rho\in\mathbb{C}^{4 \times 4}$, $\rho \geq 0$, and ${\rm Tr}\rho = 1$.

Following~\cite{Pechen_Rabitz_2006, Morzhin_Pechen_LJM_2021}, consider the GKSL master equation
\begin{equation}
\frac{d \rho(t)}{dt} = 
-i \left[H_S + \varepsilon H_{{\rm eff},n(t)} + V_{u(t)}, \rho(t) \right] + 
\varepsilon \mathcal{L}_{n(t)}(\rho(t)),\qquad \rho(t=0) = \rho_0.
\label{GKSL_eq}
\end{equation}
Here $H_S$ is the free Hamiltonian for the two qubits, $H_{{\rm eff},n(t)}$ 
is the two-qubit effective Hamiltonian (Lamb shift), which
depends on the incoherent control $n = \left( n_{\omega_1}, n_{\omega_2} \right)$, 
which, in general, is considered as a pair piecewise continuous functions, and interaction Hamiltonian $V_{u(t)} = V u(t)$ depends on real valued coherent control~$u$, which, in general, is considered as piecewise continuous function 
(physically, e.g., shaped laser field). Here $V$ is the operator 
describing the interaction between the system and the coherent field,
$\mathcal{L}_{n(t)}(\rho(t))$ is the superoperator of dissipation which depends on~$n$, 
the parameter $\varepsilon > 0$ describes strength of the coupling between 
the system and its environment, $\rho_0$ is a given initial density matrix 
corresponding to some pure or mixed quantum state. The notation $[A, B] 
= AB - BA$ denotes commutator of operators $A$ and $B$. In this article, we consider the rational system 
of units where the reduced Planck's constant~$\hbar$ and the speed of light~$c$ are equal to~1.

We consider the situation when the qubits have sufficiently different frequencies and can 
be addressed independently by incoherent control.
Free Hamiltonian $H_S$ and effective Hamiltonian $H_{\rm eff}$ are the following:
\begin{eqnarray}
H_S &=& H_{S,1} + H_{S,2} = \frac{\omega_1}{2} \left( \sigma^z \otimes \mathbb{I}_2 \right) 
+ \frac{\omega_2}{2} \left( \mathbb{I}_2 \otimes \sigma^z \right),
\label{H_S} \\
H_{{\rm eff},n(t)} &=& H_{{\rm eff},n(t),1} + H_{{\rm eff},n(t),2} =
\Lambda_1 n_{\omega_1}(t) \left( \sigma^z \otimes \mathbb{I}_2 \right) + 
\Lambda_2 n_{\omega_2}(t) \left( \mathbb{I}_2 \otimes \sigma^z \right),
\label{H_eff}
\end{eqnarray}
where $\sigma^z = \begin{pmatrix}
1 & 0 \\
0 & -1
\end{pmatrix}$
is $Z$ Pauli matrix; $\mathbb{I}_2$ is the $2 \times 2$ identity matrix;  
tensor products are $\sigma^z \otimes \mathbb{I}_2 = 
\begin{pmatrix}
\mathbb{I}_2 & 0_2 \\
0_2 & -\mathbb{I}_2
\end{pmatrix}$ and
$\mathbb{I}_2 \otimes  \sigma^z = 
\begin{pmatrix}
\sigma^z & 0_2 \\
0_2 & \sigma^z
\end{pmatrix}$; $0_2$ means the $2 \times 2$ zero matrix. Incoherent controls 
$n_1(t)\equiv n_{\omega_1}(t)$ and $n_2(t)\equiv n_{\omega_2}(t)$ are arbitrary functions 
of time defined on the interval $t \in [0, T]$, they represent density of particles 
of the environment at frequencies $\omega_1$ and $\omega_2$ and can be controlled 
independently. Because incoherent control by its physical 
meaning is a density of particles, it is a non-negative 
function of time and we have the constraints
\begin{eqnarray}
n_1(t) \geq 0, \qquad n_2(t) \geq 0 \qquad \text{for all} \quad t \in [0, T].
\label{lower_constraints_to_incoherent_controls}
\end{eqnarray}

The interaction operator $V_{u(t)} = V u(t)$ is defined, in general, with some 
arbitrary Hermitian matrix~$V$ and, in particular, as in~\cite{Morzhin_Pechen_LJM_2021}, 
we consider the following two types:
\begin{eqnarray}  
V = V_1 &:=& \sigma^x \otimes \mathbb{I}_2 + \mathbb{I}_2 \otimes \sigma^x = 
\begin{pmatrix}
0_2 & \mathbb{I}_2 \\
\mathbb{I}_2 & 0_2
\end{pmatrix} + 
\begin{pmatrix}
\sigma^x & 0_2 \\
0_2 & \sigma^x
\end{pmatrix} =
\begin{pmatrix}
\sigma^x & \mathbb{I}_2 \\
\mathbb{I}_2 & \sigma^x
\end{pmatrix},
\label{V_variant1} \\ 
V = V_2 &:=& \sigma^x \otimes \sigma^x = \begin{pmatrix}
0_2 & \sigma^x \\
\sigma^x & 0_2
\end{pmatrix},
\label{V_variant2}
\end{eqnarray} 
where $\sigma^x = \begin{pmatrix}
0 & 1 \\
1 & 0
\end{pmatrix}$ 
is $X$ Pauli matrix. The difference between these two interactions~$V$ is that 
in the case of~(\ref{V_variant1}) the same coherent control~$u$ addresses each 
qubit independently, while in the case of~(\ref{V_variant2}) the control~$u$ 
acts to couple the qubits. 

The superoperator of dissipation acts on the density matrix as
\begin{eqnarray}
\mathcal{L}_{n(t)}(\rho(t)) &=& 
\mathcal{L}_{n(t),1}(\rho(t)) + 
\mathcal{L}_{n(t),2}(\rho(t)), 
\label{dissipator}
\\
\mathcal{L}_{n(t),j}(\rho(t)) &=&  \Omega_j (n_{\omega_j}(t) + 1) \left( 2 \sigma^-_j \rho \sigma^+_j - 
\sigma_j^+ \sigma_j^- \rho - \rho \sigma_j^ + \sigma_j^- \right) + \nonumber \\
&& + \Omega_j n_{\omega_j}(t) \left( 2\sigma^+_j \rho \sigma^-_j - 
\sigma_j^- \sigma_j^+ \rho - \rho \sigma_j^- \sigma_j^+ \right),
\qquad j = 1,2, 
\label{dissipator_j}
\end{eqnarray}
where $\Lambda_j>0$ and $\Omega_j>0$ are some constants and matrices $\sigma_j^\pm$ are
\begin{equation}
\label{sigma_pm_j}
\sigma_1^{\pm} = \sigma^{\pm} \otimes \mathbb{I}_2, \qquad 
\sigma_2^{\pm} = \mathbb{I}_2 \otimes \sigma^{\pm} \qquad\text{with}\quad 
\sigma^+ = \begin{pmatrix}
0 & 0 \\ 1 & 0
\end{pmatrix}, \quad
\sigma^- = \begin{pmatrix}
0 & 1 \\ 0 & 0
\end{pmatrix}.
\end{equation}

There are various approaches for representation of density matrices, e.g., 
ordinary Bloch parametrization for density matrices of a two-level quantum system 
(e.g.,~\cite{Holevo_book_2019}), generalized Bloch vector (e.g.,~\cite{Basilewitsch_Koch_Reich_2019}), 
probability representation~\cite{Manko_Manko_Entropy_2021},~etc. Bloch parameterization in general case of $N$-level system is parameterization in some traceless Hermitian $N\times N$ matrix basis, e.g. in the basis of generalised Gell-Mann matrices~\cite{Bertlmann_Krammer_2008}. Probability representation is a recent approach which involves constructing a map from quantum state (density operator) to specific classical probability distribution. We use linear parametrization which in our approach might be natural for computing gradient of the objective and for the subsequent use of optimization tools. Use of direct matrix representation might be convenient, as was done with genetic algorithm in~\cite{Pechen_Rabitz_2006}.

Following~\cite{Morzhin_Pechen_LJM_2021}, 
the computational approach reduces the system (\ref{GKSL_eq}) including 
(\ref{H_S}), (\ref{H_eff}), (\ref{dissipator}), (\ref{dissipator_j}), and \eqref{sigma_pm_j} 
with interaction operator~$V$ defined either by~(\ref{V_variant1}) or by~(\ref{V_variant2}) 
to the corresponding form in terms of real states by considering real and imaginary 
parts of the matrix elements of density matrix~$\rho$. Taking into account 
the Hermiticity of the density matrix, denote
\begin{equation}
\rho = \begin{pmatrix}
\rho_{1,1} & \rho_{1,2} & \rho_{1,3} & \rho_{1,4} \\
\rho_{1,2}^{\ast} & \rho_{2,2} & \rho_{2,3} & \rho_{2,4} \\
\rho_{1,3}^{\ast} & \rho_{2,3}^{\ast} & \rho_{3,3} & \rho_{3,4} \\
\rho_{1,4}^{\ast} & \rho_{2,4}^{\ast} & \rho_{3,4}^{\ast} & \rho_{4,4} 
\end{pmatrix} 
= \begin{pmatrix}
x_1 & x_2 + i x_3 & x_4 + i x_5 & x_6 + i x_7 \\
x_2 - i x_3 & x_8 & x_9 + i x_{10} & x_{11} + i x_{12} \\
x_4 - i x_5 & x_9 - i x_{10} & x_{13} & x_{14} + i x_{15} \\
x_6 - i x_7 & x_{11} - i x_{12} & x_{14} - i x_{15} & x_{16} 
\label{rho_parametrization}
\end{pmatrix},
\end{equation}
where $x_j \in \mathbb{R}$, $j=1,2,\dotsc, 16$. Used in the definition of density 
matrix condition ${\rm Tr}\rho=1$ implies linear constraint 
\begin{eqnarray}
x_1+x_8+x_{13}+x_{16}=1. \label{trace_condition}
\end{eqnarray}

For the two types of~$V$, we have two different dynamical systems 
written in~\cite{Morzhin_Pechen_LJM_2021} with 16-dimensional real-valued state~$x$. 
Both these systems belong to the following general class of bilinear homogeneous systems:
\begin{equation}
\label{dynamical_system_x_common_form}
\frac{dx}{dt} = \left(A + B_u u + B_{n_1} n_1 + B_{n_2} n_2 \right) x, \qquad x(0) = x_0, 
\end{equation}  
where the $16 \times 16$ matrices $A$, $B_u$, $B_{n_1}$, $B_{n_2}$ are found after 
substituting the parameterization~(\ref{rho_parametrization}) 
in the GKSL equation~(\ref{GKSL_eq}); $x_0$ is found from a given $\rho_0$. 

For both types of~$V$, the explicit forms for all the 16~differential equations and the corresponding 
initial conditions in (\ref{dynamical_system_x_common_form}) were obtained  in~\cite{Morzhin_Pechen_LJM_2021}. 
For brevity, we do not reproduce here these 32~differential equations and the corresponding 
$16 \times 16$ matrices.

We denote the full control $c = (u, n_1, n_2)$ and consider the following objective functional to be minimized  
for a given target density matrix~$\rho_{\rm target}$ and a given final time~$T$ --- it describes the problem of obtaining $\rho(T)$ being as close as possible to~$\rho_{\rm target}$ 
in the Hilbert--Schmidt distance:
\begin{eqnarray}
J_{\rm dist.}(c) = F_{\rm dist.}(\rho(T); \rho_{\rm target}) := \| \rho(T) - \rho_{\rm target} \|^2 \to \inf. 
\label{problem_minimize_distance}
\end{eqnarray}

In terms of the parameterization~(\ref{rho_parametrization}), the problem~(\ref{problem_minimize_distance}) 
is reformulated as the following~\cite{Morzhin_Pechen_LJM_2021}:  
\begin{equation}
J_{\rm dist.}(c) = \mathcal{F}_{\rm dist.}(x(T);x_{\rm target}) := \left\langle x(T), Z x(T) \right\rangle + 
\left\langle b, x(T) \right\rangle + d \to \inf,
\label{problem_minimize_distance_in_terms_of_x}
\end{equation}
where $Z = {\rm diag}(\beta)$, $b = -2 \beta \odot x_{\rm target}$, 
$d = \left\langle \beta \odot x_{\rm target}, x_{\rm target} \right\rangle$, 
$\beta =$ (1, 2, 2, 2, 2, 2, 2, 1, 2, 2, 2, 2, 1, 2, 2, 1),   
and ``$\odot$'' denotes the Hadamard product. Thus, the terminal function 
$\mathcal{F}_{\rm dist.}(x;x_{\rm target})$ is linear-quadratic and convex. 
Here $\beta_j = 1$, if $j \in \{1, 8, 13, 16\}$ that is related to those components of~$x$ 
which are on the main diagonal in~(\ref{rho_parametrization}). At $x = x_{\rm target}$, 
the function  $\mathcal{F}_{\rm dist.}(x, x_{\rm target})$ has zero value.

The problem (\ref{problem_minimize_distance}) can be used
as an auxiliary problem with some $T$ for the time-minimal 
steering problem $\rho_0 \to \rho_{\rm target}$ (for an one-qubit case,
this approach was used in~\cite{Morzhin_Pechen_IJTP_2021_2019}).

In this article, we consider variable in time coherent and incoherent controls, which, in general,
are piecewise continuous functions or, in particular, piecewise constant controls 
\begin{eqnarray}
	u(t) &=& \sum\limits_{j=1}^N u^j \chi_{[t_{j-1}, t_j)}(t), 
	\label{pc_coherent_control} \\
	n_i(t) &=& \sum\limits_{j=1}^N n_i^j \chi_{[t_{j-1}, t_j)}(t), 
	\qquad i = 1,2 
	\label{pc_incoherent_control}
\end{eqnarray} 
where $0 = t_0 < t_1 < \dots < t_N = T$ and $\chi_{[t_{j-1}, t_j)}$ 
is the characteristic function of $[t_{j-1}, t_j)$.

When piecewise constant controls (\ref{pc_coherent_control}) and 
(\ref{pc_incoherent_control}) are used, we combine all the variables defining such 
controls $u$,~$n_1$,~and~$n_2$ in one vector 
\begin{eqnarray}
{\bf a} = (a_1, a_2, \dots, a_{3N}) := \big(u^1, u^2, \dots, u^N, 
n_1^1, n_1^2, \dots, n_1^N, n_2^1, n_2^2, \dots, n_2^N \big).
\label{vector_parameters_defining_controls}
\end{eqnarray}
The objective functional $J_{\rm dist.}(c)$ then becomes  
the objective function: 
\begin{eqnarray*}
g_{\rm dist.}({\bf a}) &:=& F_{\rm dist.}(\rho(T; {\bf a}); \rho_{\rm target}) = \mathcal{F}_{\rm dist.}(x(T; {\bf a}); x_{\rm target}) \to \inf, 
\label{objective_function_g_dist} 
\end{eqnarray*}
where $\rho(\cdot; {\bf a})$ and $x(\cdot; {\bf a})$ are the solutions, 
correspondingly, of the systems (\ref{GKSL_eq}) and (\ref{dynamical_system_x_common_form})
for piecewise constant controls corresponding to some admissible~${\bf a}$. 

In post-optimization analysis for the problem (\ref{problem_minimize_distance}) 
with optimized control $c = (u, n_1, n_2)$ and the corresponding solution $\rho$
of the system (\ref{GKSL_eq}), one can study, in addition,
how the values of $F_{\rm dist.}(\rho(t); \rho_{\rm target})$ are changed
when $t$ goes from $t=0$ to $t=T$ for a given target 
density matrix $\rho_{\rm target}$. 
For the same problem,  (\ref{problem_minimize_distance}), one can also analyze the behaviour of such quantities
characterizing the two-qubit system as von Neumann entropy and purity versus time. 
They are defined as functions of density matrix $\rho$ as:
\begin{itemize}
\item {\it von Neumann entropy} \cite{Holevo_book_2019}
\begin{eqnarray}
    S(\rho) = - \mathrm{Tr} \left(\rho \log_e \rho \right) = 
    -\sum_{\lambda_i \neq 0} \lambda_i \log_e \lambda_i\in[0,\log_e \mathrm{dim} \mathcal{H}],
    \label{von_Neumann_entropy}
\end{eqnarray}
\item {\it purity}
\begin{eqnarray}
    P(\rho) = \mathrm{Tr} \rho^2 = \langle \rho, \rho \rangle =  \sum_{i,j} |\rho_{ij}|^2 \in \left[\frac{1}{\dim \mathcal{H}}, 1 \right],
\label{purity}
\end{eqnarray} 
\end{itemize}
where $\lambda_i$ are eigenvalues of the density matrix $\rho$ and $\dim\mathcal{H}$ 
is the dimension of the Hilbert space, which is four in our case. As it is known, entropy 
is minimal for pure states for which it equals to zero, i.e. if 
$\rho_\mathrm{pure} = |\psi\rangle\langle\psi|$ is a pure state, then $S(\rho_\mathrm{pure}) = 0$. 
Maximum value of entropy is obtained for completely mixed quantum state 
$\rho = \mathbb{I} / \mathrm{dim} \mathcal{H}$ and equals $S_\mathrm{max} = \log_e \mathrm{dim} \mathcal{H}$. 
For the considered 4-dimensional system (\ref{GKSL_eq}) we have 
$S_\mathrm{max} = S(\mathbb{I} / 4)=\log_e 4 \approx 1.386$. Purity is the main quantity 
that characterizes how close the system state is to a pure state. It attains maximum value $P_{\rm max} = 1$ 
at pure states and minimum value $P_{\rm min} = 1 / \mathrm{dim} \mathcal{H}$ at the completely 
mixed state. In terms of the vector $x$, purity is
$$P = \sum_{i,j} |\rho_{ij}|^2 = \langle \beta \odot x, x \rangle,$$
where $\beta$ is defined after Eq.~(\ref{problem_minimize_distance_in_terms_of_x}). 

For the formulated above optimization problem~(\ref{problem_minimize_distance}), 
we also study the behavior of each qubit individually. Corresponding reduced density 
matrices $\rho^i \in  \mathbb{C}^{2\times2},\; i=1,2,$ are defined via partial trace as:
\begin{equation}
\rho^1 = \mathrm{Tr}_{{\cal H}_2} \rho = \sum_{k = 1}^2 (\mathbb{I} \otimes \langle k |) \rho (\mathbb{I} \otimes | k\rangle ),
\label{reduced_rho_1}
\end{equation}
\begin{equation}\rho^2 = \mathrm{Tr}_{{\cal H}_1} \rho = \sum_{k = 1}^2 (\langle k |  \otimes \mathbb{I}) \rho (|k\rangle \otimes  \mathbb{I}),
\label{reduced_rho_2}
\end{equation}
where $\mathbb{I}$ is identity operator on $\mathcal{H}_i$,  $| k\rangle$ are basis vectors in $\mathcal{H}_i$, and ''$\otimes$'' denotes tensor product. Density matrix of a qubit can be 
bijectively mapped to the Bloch ball (ball in $\mathbb{R}^3$ with \mbox{radius 1)} via the following parameterization:
$$\quad r_j = \mathrm{Tr}\rho \sigma_j,\quad j \in \{x, y, z\}, \quad \sigma = (\sigma_x, \sigma_y, \sigma_z),$$
where $\sigma_j$ are Pauli matrices and Bloch vector  $r=(r_x,r_y,r_z)$ satisfies  $|r|\le 1$. 
In terms of vector $x$, Bloch vectors of the first and second qubits are
\begin{equation}
    r^1 = \big(
    2 (x_4 + x_{11}),\;
    -2 (x_5 + x_{12}),\;
    x_1 + x_8 - x_{13} - x_{16}\big),
    \label{reduced_bloch_1}
\end{equation}
\begin{equation}
    r^2 = \big(
    2 (x_2 + x_{14}),\;
    -2 (x_3 + x_{15}),\;
    x_1 + x_{13} - x_8 - x_{16}
    \big).
    \label{reduced_bloch_2}
\end{equation}

The optimization problem~(\ref{problem_minimize_distance}) was formulated for a general quantum system 
in~\cite{Pechen_Rabitz_2006}, where two-qubit case with time-independent incoherent controls was also 
studied. One-qubit system driven by piecewise continuous coherent and incoherent controls was studied 
in more details in~\cite{Morzhin_Pechen_IJTP_2021_2019}, where the problem of minimizing the 
Hilbert--Schmidt distance's square for a fixed final time was studied as an auxiliary problem for solving 
a control problem of steering $\rho_0 \to \rho_{\rm target}$ in a minimal possible time. For the same 
one-qubit system, the article~\cite{Morzhin_Pechen_LJM_2020} considered also time-minimal control problem, 
but with piecewise constant coherent and incoherent controls together with the requirements to satisfy 
the terminal constraint $\rho(T) = \rho_{\rm target}$ and minimize the final time~$T$. Such class of controls 
was used in~\cite{Morzhin_Pechen_LJM_2020} for considering these parameters together with~$T$ as outputs in 
the regression problem for obtaining suboptimal solutions of the time-minimal problem; here certain machine 
learning techniques were used. In~\cite{Morzhin_Pechen_AIP_Conf_Proc_2021}, the problem of minimizing 
the Hilbert--Schmidt distance with a fixed final time was used for numerical estimation of reachable and 
controllability sets of a one-qubit system in the Bloch ball.
In \cite{Morzhin_Pechen_Physics_of_Particles_and_Nuclei_2020}, the Uhlmann--Jozsa fidelity of 
the final density matrix, $\rho(T)$, for the one-qubit system driven by piecewise continuous coherent 
and incoherent controls was studied. 

\section{Adopting Gradient Ascent Pulse Engineering Approach}\label{GRAPE}

\subsection{Exact Formula for the Gradient of the Objective Function}

In this section, we adopt the general idea of the GRAPE method for finding optimal shape of control for the problem~ (\ref{problem_minimize_distance}). As a first step, we reduce the initial problem to a finite-dimensional optimization problem with piecewise constant control. Then we compute an analytical expression for the gradient which is then used for a  gradient-based numerical optimization method; in this work we use first-order gradient descent method. The main advantage of this approach comes from the ability to compute analytical expression for the gradient.

In the considered optimization problem, the system (\ref{dynamical_system_x_common_form}) is driven by coherent and incoherent controls. Therefore implementing GRAPE method faces the obstacle: incoherent control is bounded below by zero, so in the control space we have a boundary which is undesirable for ordinary gradient methods. Here we analyse the unconstrained case. For that let us make a change to other control variables $w_i(t)$ with values in $\mathbb R$ that are not constrained, via the relation
$$n_i(t) = w_i^2(t), \qquad i=1,2, \qquad t \in [0, T].$$

We approximate $u$, $n_1$, and $n_2$ by piecewise constant functions (\ref{pc_coherent_control}, \ref{pc_incoherent_control}). For unconstrained optimization, we introduce piecewise constant $w_i\in\mathbb R$ and define 
\begin{equation}
	n_i^j = (w_i^j)^2, \qquad i = 1,2, \qquad j = 1,\dots,N.
	\label{GRAPE_pc_incoherent}
\end{equation}
Considering $v = (u, w_1, w_2)$ as control  we can implement GRAPE for the optimization problem (\ref{problem_minimize_distance}).
After this piecewise constant approximation, the objective functional (\ref{problem_minimize_distance_in_terms_of_x}) $J_{\rm dist.}(c)$ becomes a function of $3N$ variables that can be optimized by finite-dimensional optimization methods.

Evolution of the system (\ref{dynamical_system_x_common_form}) is composition of matrix exponentials: 
\begin{equation}
	x(T) = e^{\Delta t_N L_N} \cdots e^{\Delta t_1 L_1} x_0,
	\label{GRAPE_evolution}
\end{equation}
where $\Delta t_j = t_j - t_{j-1}$ and $L_j$ is the right hand side matrix of the equation (\ref{dynamical_system_x_common_form}) at the moment $t \in [t_{j-1}, t_j),\; j = 1, \dots, N$:
$$L_j = A + B_u u^j + B_{n_1} (w_1^j)^2 + B_{n_2} (w_2^j)^2.$$

Gradient of the functional (\ref{problem_minimize_distance_in_terms_of_x}) with respect to control $v = (u, w_1, w_2)$ can be computed via the chain rule:
\begin{equation}
	\dfrac{\delta J_{\rm dist.}}{\delta v} = \dfrac{\delta \mathcal{F}_{\rm dist.}}{\delta x(T)} \dfrac{\delta x(T)}{\delta v}.
	\label{GRAPE_gradient_J_v}
\end{equation}
Differentiating (\ref{problem_minimize_distance_in_terms_of_x}) gives us
\begin{equation*}
	\dfrac{\delta \mathcal{F}_{\rm dist.}}{\delta x(T)} = 2 Z x(T) + b.
\end{equation*}
Thus gradient of $x(T)$ with respect to the control $v = (u, w_1, w_2)$ remains to be found. If the control is piecewise constant (\ref{pc_coherent_control}, \ref{GRAPE_pc_incoherent}), then partial derivatives of $x(T)$ (\ref{GRAPE_evolution}) with respect to $v^j = (u^j, w_1^j, w_2^j)$ are computed as
\begin{equation}
	\dfrac{\partial x(T)}{\partial v^j} = e^{\Delta t_N L_N} \cdots e^{\Delta t_{j + 1} L_{j + 1}} \dfrac{\mathrm{d}}{\mathrm{d} v^j} \left(e^{\Delta t_{j} L_{j}}\right) e^{\Delta t_{j - 1} L_{j - 1}} \cdots e^{\Delta t_1 L_1} x_0.
	\label{GRAPE_exact_1}
\end{equation}
Since $A$, $B_u$, $B_{n_1}$, $B_{n_2}$ do not commute with each other, we have to use the following special integral formula \cite{Wilcox1967} for derivative of matrix exponential:
\begin{equation}
	\dfrac{\mathrm{d}}{\mathrm{d} v^j} e^{\Delta t_{j} L_{j}} = \Delta t_j \int_0^1  \,\exp(\alpha \Delta t_j L_j) \dfrac{\mathrm{d} L_j}{\mathrm{d} v^j}\exp((1 - \alpha) \Delta t_j L_j)\mathrm{d}\alpha,
	\label{GRAPE_exact_2}
\end{equation}
where $\dfrac{\mathrm{d} L_j}{\mathrm{d} v^j}$ have different forms for coherent and incoherent components of $v$:
\begin{equation}
	\dfrac{\partial L_j}{\partial u^j} = B_u,\qquad \dfrac{\partial L_j}{\partial w_i^j} = 2w_i^jB_{n_i},\qquad i = 1,2.
	\label{GRAPE_exact_3}
\end{equation}
Now after obtaining the exact formula (\ref{GRAPE_exact_1}) -- (\ref{GRAPE_exact_3}) for gradient of the functional (\ref{problem_minimize_distance_in_terms_of_x}) with respect to piecewise constant control (\ref{pc_coherent_control}, \ref{GRAPE_pc_incoherent}), gradient search can be implemented for numerical solving of the optimization problem~(\ref{problem_minimize_distance}).

\subsection{Numerical Results}
\label{GRAPE_numerical_results}

Here, a numerical simulation of GRAPE algorithm for the state-to-state transfer optimization problem (\ref{problem_minimize_distance}) is performed to demonstrate the abilities for using gradient search in solving problems of generating target states using coherent and incoherent controls in two-qubit systems. For numerical simulation we consider the both types of the interaction operator $V$, i.e.~(\ref{V_variant1}) and~(\ref{V_variant2}), and use the following values of the system's parameters: $\varepsilon = 0.1$, $\omega_1 = 1$, $\omega_2 = 0.5$, $\Lambda_1 = \Lambda_2 = 0.05$, and $\Omega_1 = \Omega_2 = 0.05$. For the state-to-state transfer problem choice of the two parameters is important: final time $T$ and number of partition intervals $N$ in (\ref{pc_coherent_control}) and (\ref{pc_incoherent_control}). Changing the final time $T$ can influence the ability of steering the system to the target state. Generally one can expect that decreasing the final time $T$ can lead to smaller degree of controllability of the system. In opposite, increasing of $N$ obviously gives more freedom for controlling the system. We consider here the final time $T = 5$ and regular partition of the segment $[0, T]$ into $N = 10$ time intervals $\Delta t_j$, so that each $\Delta t_j = T/N = 0.5$. For the initial and the target states we choose $\rho_0 = {\rm diag}(0.9, 0.1, 0, 0)$ and $ \rho_\mathrm{target} = {\rm diag}(0.2, 0.3, 0.2, 0.3)$, that in terms of $x$ are $x_0 =  (0.9, \text{six zeros}, 0.1, \text{eight zeros})$ and 
$x_{\rm target} = (0.2, \text{six zeros}, 0.3, \text{four zeros}, 0.2, 0, 0, 0.3)$.

For numerical finite-dimensional optimization method we chose ordinary gradient descent (similarly gradient ascent can be used for maximization). This is a first-order iterative algorithm for finding local minimum (descent) or maximum (ascent) of differentiable function. In terms of the optimization problem (\ref{problem_minimize_distance_in_terms_of_x}), iterative formula for $(k + 1)$-th step of the algorithm can be written as follows:
\begin{equation}
	v^{(k + 1)} = v^{(k)} - h_k \mathrm{grad}_v J(v^{(k)}), \qquad k = 0, 1, \dots
	\label{GRAPE_gradient_descent}
\end{equation}
where 
$$\mathrm{grad}_v = \left(\dfrac{\partial}{\partial v^1}, \dots , \dfrac{\partial}{\partial v^N} \right).$$ 
Thus \,$\mathrm{grad}_v J(v^{(k)})$\, is equal to (\ref{GRAPE_gradient_J_v}) with $v = v^{(k)}$; $h_k$ are the values of the iterations steps. In Eq.~(\ref{GRAPE_gradient_descent}), the notation $v$ means that this formula is written for each of three components of $v = (u, w_1, w_2)$.

\begin{figure}[ht!]
	\centering
	\includegraphics[width = 1\linewidth]{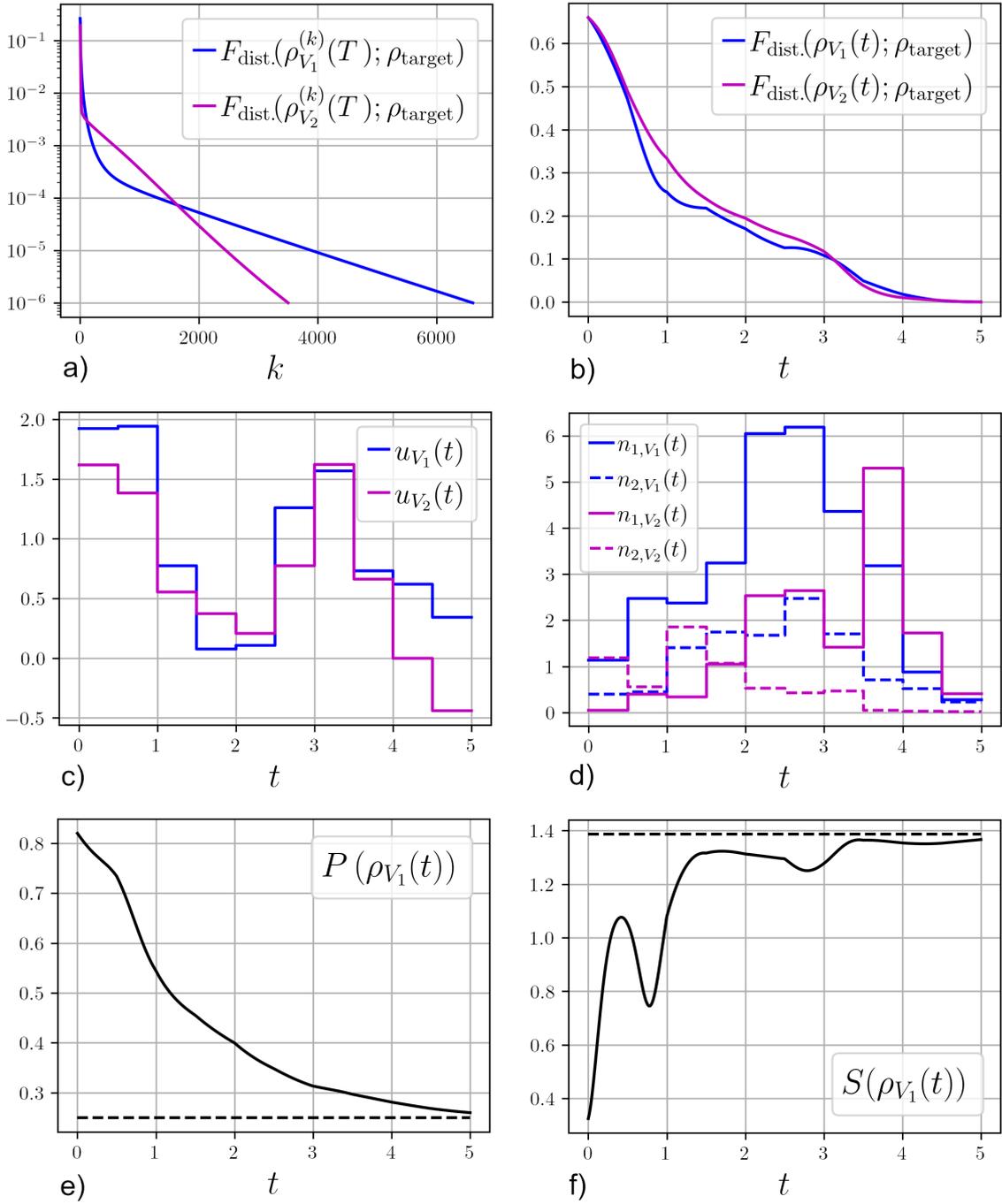}
	\caption{\label{Fig1} For Subsection \ref{GRAPE_numerical_results}. 
    Optimal coherent $u(t)$ (1c) and incoherent $(n_1(t), n_2(t))$ (1d) controls for the problem of state-to-state transfer, i.e. minimizing of the functional (\ref{problem_minimize_distance_in_terms_of_x}) for two types of interaction operator $V = V_1$ (blue) and $V = V_2$ (purple). 
    Convergence of $\mathcal{J}_\mathrm{dist.}(v^{(k)}) = F_\mathrm{dist.} ( \rho^{(k)}(T);\rho_\mathrm{target})$ to zero with iterations is shown on subplot~1a. Other subplots show dynamics of Hilbert-Schmidt distance $F_\mathrm{dist.} \left( \rho(t);\rho_\mathrm{target}\right)$ (1b), von Neumann entropy $S(\rho(t))$  (1f) and purity $P(\rho(t))$ (1e) of the two-qubit density matrix $\rho(t)$ of the system evolving  under optimal control shown on (1c, 1d). Dotted lines on subpots (1f) and (1e) show, respectively, maximal value of entropy 
    (entropy of the completely mixed state)
    , which is $\log_e 4$, and minimal value of purity
    (purity of the completely mixed state)
    , which equals to~$1/4$.}
\end{figure}

For gradient descent method (\ref{GRAPE_gradient_descent}) we chose initial guess for the control $v^{(0)} = (u^{(0)}, w_1^{(0)}, w_2^{(0)})$ as ${u^{(0)}}^j = \cos(0.3 t_j)$ and ${w_1^{(0)}}^j = {w_2^{(0)}}^j = 
\exp\left(-5 \left(t_j/T - 1/2 \right)^2 \right)$, $j = 1,2,\dots,N$. Iterations of the gradient descent stop when the following stopping criterion is satisfied: 
\begin{equation}
	J_\mathrm{dist.}(\bar{u}, \bar{w}_1, \bar{w}_2) = \mathcal{F}_\mathrm{dist} \left( x(T);x_\mathrm{target}\right)  < \epsilon,
	\label{GRAPE_stop_criterion}
\end{equation}
thus we find control $\bar{v} = (\bar{u}, \bar{w}_1, \bar{w}_2)$ which steer the system to the final state $x(T)$ that almost equals $x_\mathrm{target}$, i.e. differs by not more than accuracy $\epsilon$. Gradient descent over trap-free quantum control landscapes can generally be faster than global optimization methods. In presence of traps, global search methods would generally be more preferable. While the structure of the landscape is not know for the considered control problem, we set a small accuracy $\epsilon = 10^{-6}$. Finally, we set constant value of steps $h_k = h = 1$.

Figure~1a shows the behaviour of $F_\mathrm{dist.} ( \rho^{(k)}(T);\rho_\mathrm{target})$ over the first 200 iterations for the first type of interaction operator $V = V_1$ (blue lines) and second type of interaction operator $V = V_2$ (purple line), where $\rho^{(k)}(t)$ is the density matrix of the system for $k$th approximation~(\ref{GRAPE_gradient_descent}) of control $v = (u, w_1, w_2)$. Overall, it took $\approx 6600$ iterations for $V = V_1$ and $\approx 3500$ iterations for $V = V_2$ to reach the accuracy $\epsilon = 10^{-6}$.  Comparing two different types for the interaction operator $V$, it may be inferred that for the chosen parameters the
algorithm converges faster in the case $V = V_2$. In the context of the problem of steering a given initial state $\rho_0$ to a specific target state $\rho_\mathrm{target}$~(\ref{problem_minimize_distance}), this means that if this kind of difference remains for other parameters then the second type of interaction operator $V = V_2$~(\ref{V_variant2}) can be more preferable.

Figures 1c and 1d show the optimal coherent $u(t)$ and incoherent $n(t) = (n_1(t), n_2(t))$ controls for two types of the interaction operator $V$~(\ref{V_variant1}) and (\ref{V_variant2}), which were found numerically with accuracy $\epsilon = 10^{-6}$  (\ref{GRAPE_stop_criterion}).  

Figure 1b shows the dynamics of the Hilbert-Schmidt distance between $\rho(t)$ and $\rho_\mathrm{target}$, i.e. values of the following functional depending on time $t \in [0, T]$:
\begin{equation}
    F_\mathrm{dist} \left( \rho(t);\rho_\mathrm{target}\right) = \|\rho(t) - \rho_\mathrm{target}\|^2.
    \label{GRAPE_HS_dist}
\end{equation}
This distance decreases with time $t$, starting from some value at $t = 0$ and tends to almost zero at $t = T$, when it coincides with the value of the optimized functional~(\ref{GRAPE_stop_criterion}).

Figure 1f shows evolution of the von Neumann entropy for $V = V_1$, which is defined by~(\ref{von_Neumann_entropy}), of the density matrix $\rho(t)$ evolving under the controls obtained after the optimization. Figure 1f shows that entropy has the value of $S \approx 0.3$ at $t = 0$, then increases with some fluctuations with increasing  time $t$ and ends at the higher value $S \approx 1.366$ at $t = T$ that is very close to the value of completely mixed state $S_{\rm max} = 1.386$ (dashed line in figure 1f). This is because the system starts at the state $\rho_0 = {\rm diag}(0.9, 0.1, 0, 0)$ that is close to the pure state $\rho = {\rm diag}(1, 0, 0, 0)$ (ground state) and reaches the target state $\rho_\mathrm{target} = {\rm diag}(0.2, 0.3, 0.2, 0.3)$ that is close to completely mixed state $\rho = {\rm diag}(0.25, 0.25, 0.25, 0.25)$.

The main quantity that characterizes how close the system is state to pure states, is purity defined by Eq.~(\ref{purity}). Figure 1e shows evolution of purity $P(\rho(t))$ of the state $\rho(t)$ for the two-qubit system with first type of the interaction operator $V = V_1$ evolving under the controls obtained after the optimization. The system in the initial state $\rho(0) = {\rm diag}
(0.9, 0.1, 0, 0)$ at $t = 0$ has purity $P \approx 0.82$ relatively close to maximum $P_{\rm max} = 1$, then purity decreases with increasing time $t$ and approaches the final value $\gamma \approx 0.26$ at $t = T$, which is close to the minimal value of purity $P_\mathrm{min} = 1/4 = 0.25$ (which is shown by dashed line in figure 1e).

\begin{figure}[ht!]
	\centering
	\includegraphics[width = 1\linewidth]{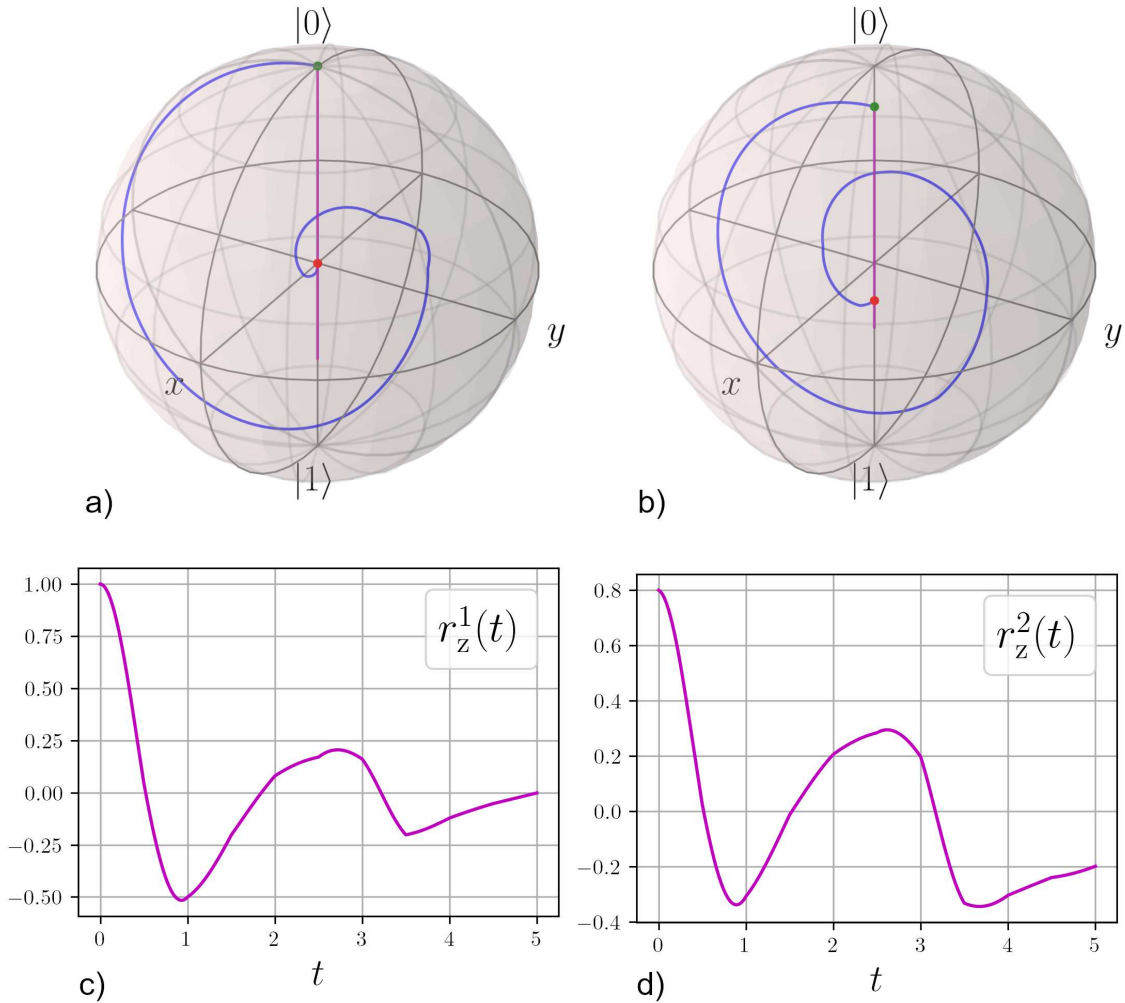}
	\caption{Evolution of the Bloch vectors of the reduced density matrices for the first (subplot (a)) and second (subplot (b)) qubit under optimal control (Figs. 1(c) and 1(d)) for first (blue) and second (purple) type of interaction $V$. The parameters are the same as on Fig.~\ref{Fig1}. Green points show Bloch vectors of the initial states and red points show Bloch vectors of the target states for each qubit. Subplots~2c~and~2d reveal dynamics of $z$-coordinate of Bloch vectors of the first (left) and second (right) qubit.}
\end{figure}

Finally, figures 2a and 2b show the dynamics of the two qubits as the evolution of  Bloch vectors~(\ref{reduced_bloch_1}, \ref{reduced_bloch_2}) of their reduced density matrices ~(\ref{reduced_rho_1}, \ref{reduced_rho_2}) under the obtained optimal control (shown on figures 1c and 1d) for both types of interaction operator $V = V_1$ (blue line) and $V = V_2$ (purple line). 

The initial state $\rho_0 = {\rm diag}(0.9, 0.1, 0, 0)$ and the target state $\rho_\mathrm{target} = {\rm diag}$(0.2, 0.3, 0.2, 0.3) are separable so that they can be represented as tensor product of reduced density matrices $\rho_0 = \rho_0^1 \otimes \rho_0^2$ and $\rho_\mathrm{target} = \rho_\mathrm{target}^1 \otimes \rho^2_\mathrm{target}.$ The corresponding Bloch vectors of the initial states are $r_0^1 = |0\rangle = (0,0,1)$ (ground state) and $r_0^2 = (0, 0, 0.8)$ (green points in figure 2a, 2b), the target states are $r_\mathrm{target}^1 = (0,0,0)$ (completely mixed state) and $r_\mathrm{target}^2 = (0, 0, -0.2)$ (red points in figure 2a, 2b). 

It can be noted that trajectories of the first and second Bloch vectors for $V = V_2$ (purple lines in~2a and~2b) are straight, while trajectories in case $V = V_1$ are curved. For second type of interaction coordinates $x$ and $y$ are zeros on $[0, T]$, so figures~2c and~2d show dynamics of $z$-coordinates of first and second Bloch vectors $r_z^1(t)$ and $r_z^2(t)$.  This behavior of $x$- and $y$-coordinates of Bloch vectors in case $V = V_2$ can be explained as follows. Consider vector $\tilde{x}$ combining certain components of vector $x$ which correspond to some of the non-diagonal elements of density matrix $\rho$~(\ref{rho_parametrization}):
$$\tilde{x} = (x_2, x_3, x_4, x_5, x_{11}, x_{12}, x_{14}, x_{15}).$$
It turns out that they evolve independently on other components of vector $x$, i.e.
$$\dfrac{d \tilde{x}}{d t} = \tilde{A} \tilde{x},$$
where $\tilde{A}$ is a $8 \times 8$ matrix. If the initial state $\rho_0$ is diagonal, then $\tilde{x}(0) = 0$, so that $\tilde{x}(t) \equiv~0$ for all $t~\in~[0, T]$. Since $x$- and $y$-coordinates of Bloch vectors $r_1$~(\ref{reduced_bloch_1} and $r_2$~\ref{reduced_bloch_2}) are linear combinations of vector $\tilde{x}$ components, they also remains equals to zero if initial state is diagonal.

\begin{figure}[ht!]
	\centering
	\includegraphics[width = 1\linewidth]{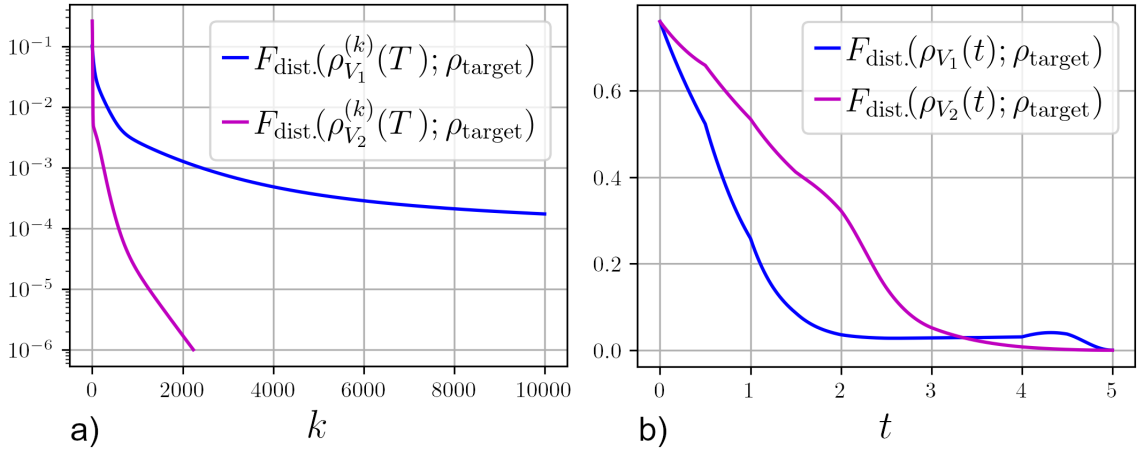}
	\caption{For Subsection \ref{GRAPE_numerical_results}. The same as on Fig.~\ref{Fig1} but for the problem of transferring the Bell state $|\Phi\rangle=(|00\rangle+|11\rangle)/\sqrt{2}$ to a separable mixed state 
		$\rho_\mathrm{target} ={\rm diag}(0.2, 0.3, 0.2, 0.3)$. While for the interaction Hamiltonian $V_2$ the algorithm converges fast, highly inefficient behaviour for the interaction Hamiltonian  $V_1$ is observed on the subplot~3a, with slow convergence and higher by several orders of magnitude obtained minimal value of the Hilbert-Schmidt distance.\label{Vadim_Fig2}}
\end{figure}

As another example, we study steering the entangled Bell state $|\Phi\rangle=(|00\rangle+|11\rangle)/\sqrt{2}$ into the separable mixed state $\rho_{\rm target}= {\rm diag}(0.2, 0.3, 0.2, 0.3)$. The results are provided on Fig~\ref{Vadim_Fig2}. In this case, the subplot~3a shows that gradient search for the model with interaction Hamiltonian $V_2$ converges, while for the interaction $V_1$ drastically different behaviour is observed with significantly slower convergence and by several orders of magnitudes higher obtained value of the objective. 

Formulae associated with gradient computation in our realization were computed via various numerical instruments (numerical methods, libraries, etc.). Almost all computations were performed using \textit{NumPy} Python library, which is very efficient in computing matrix operations. Matrix exponentials were computed via the function \texttt{scipy.linalg.expm} of \textit{SciPy} library that uses Padé's approximation. The main complexity for performed numerical simulations are related to the computation of the integral formula for matrix exponential gradient (\ref{GRAPE_exact_2}). Integral of the matrix function was computed using trapezoidal formula with error of computation equal to
$$\epsilon_\mathrm{int}^j \leq \dfrac{1}{3 N_\mathrm{int}^3}\left(\dfrac{T}{N}\right)^3 \|L_j\|^2 \left\|\frac{\mathrm{d} L_j}{\mathrm{d} v^j}\right\| \left\|\exp\left(\dfrac{T}{N}L_j\right)\right\| \lesssim \dfrac{1}{N_\mathrm{int}^3} \approx 10^{-4},$$
where $N_\mathrm{int}$ is the number of points that interpolate the function via trapezoidal rule. We used $N_\mathrm{int} = 20$ which ensures appropriate accuracy $\epsilon_\mathrm{int} \approx 10^{-4}$.
\vspace{0.5cm}

\section{Conclusions} 

In this work, we have studied a system of two qubits driven by coherent and incoherent time-dependent controls. Two physically different models of interaction with coherent control in the Hamiltonian are considered. In the first model, the same coherent control drives the qubits independently and acts as magnetic field along $x$ axis, while in the second model coherent control induces a joint dynamics of both qubits via controlled XX interaction. The decoherence term is the same for both cases and corresponds to the weak coupling model well known in theory of open quantum systems and quantum optics. Coherent control models either laser or magnetic field, while incoherent control models spectral density of incoherent photons. For this controlled system, the control problem of minimization of the Hilbert--Schmidt distance's square for the final density matrix and a given target density matrix is considered and a gradient based optimization approache is adopted, GRadient Ascent Pulse Engineering (GRAPE), which is applied to find close to optimal controls. GRAPE depends on its parameters to be adjusted in simulations. For GRAPE, taking into account that incoherent control by its physical meaning is a non-negative function of time, 
we derive an analytical expression for the gradient of the objective and develop optimization approach
based on adaptation for this problem of GRAPE strategy. Gradient computation is then reduces to matrix multiplication without the need for solving differential evolution equations. In the numerical simulations with GRAPE, steering either a mixed separable or Bell entangled state into a target mixed separable state is analyzed. We studied evolution of the von Neumann entropy, purity, reduced density matrices, and analyzed the two physically different models of interaction with coherent control in the Hamiltonian, 
for which a significantly different behavior under optimization was found. Namely, for the second model we observe significantly faster convergence of GRAPE algorithm towards minimum of the objective that indicates, based on the considered examples, that finding optimal controls in the second model appears to be relatively simpler and joint controlled XX interaction between the qubits allows for simpler finding of close to optimal controls.

\vspace{6pt} 

\section*{Acknowledgments}
This work was partially supported by the State Program of the Ministry of Science and Higher Education of
the Russian Federation (project no. 0718-2020-0025).


\begin{thebibliography}{0}

\bibitem{Glaser_Boscain_Calarco_et_al_2015} 
Glaser, S.J.; Boscain, U.; Calarco, T.; Koch, C.P.; K\"{o}ckenberger, W.; 
Kosloff, R.; Kuprov, I.; Luy, B.; Schirmer, S.; Schulte-Herbr\"{u}ggen, T.;
Sugny, D.; Wilhelm, F.K. Training Schr\"{o}dinger's cat: quantum optimal control. 
Strategic report on current status, visions and goals for research in Europe.
{\em Eur. Phys.~J.~D} {\bf 2015}, {\em 69} (12), 279, doi: \href{https://doi.org/10.1140/epjd/e2015-60464-1}{10.1140/epjd/e2015-60464-1}.

\bibitem{Koch_arXiv2022}
Koch, C.P.; Boscain, U.; Calarco, T.; 
Dirr, G.; Filipp, S.; Glaser, S.J.; Kosloff, R.; Montangero, S.; 
Schulte-Herbr\"{u}ggen, T.; Sugny, D.; Wilhelm, F.K. 
Quantum optimal control in quantum technologies. 
Strategic report on current status, visions and goals 
for research in Europe. {\em EPJ Quantum Technol.}, {\bf 2022}, {\em 9}, 19, doi: \href{https://doi.org/10.1140/epjqt/s40507-022-00138-x}{10.1140/epjqt/s40507-022-00138-x}. 

\bibitem{Pechen_Prokhorenko_Wu_Rabitz_2008}
Pechen, A.; Prokhorenko, D.; Wu, R.; Rabitz, H. 
Control landscapes for two-level open quantum systems. 
{\em J.~Phys.~A: Math. Theor.} {\bf 2008}, {\em 41}, 045205, doi: \href{https://doi.org/10.1088/1751-8113/41/4/045205}{10.1088/1751-8113/41/4/045205}.

\bibitem{Wu_Pechen_Rabitz_Hsieh_Tsou_2008}
Wu, R.; Pechen, A.; Rabitz, H.; Hsieh, M.; Tsou, B. Control landscapes for 
observable preparation with open quantum systems. {\em J.~Math. Phys.}
{\bf 2008}, {\em 49}, 022108, doi: \href{https://doi.org/10.1063/1.2883738}{10.1063/1.2883738}. 

\bibitem{Oza_Pechen_Dominy_Beltrani_Moore_Rabitz_2009}
Oza, A.; Pechen, A.; Dominy, J.; Beltrani, V.; Moore, K.; Rabitz, H.
Optimization search effort over the control landscapes for open quantum 
systems with Kraus-map evolution. {\em J.~Phys.~A: Math. Theor.}
{\bf 2009}, {\em 42}, 205305, doi: \href{https://doi.org/10.1088/1751-8113/42/20/205305}{10.1088/1751-8113/42/20/205305}.

\bibitem{Pechen_Rabitz_2006} 
Pechen, A.; Rabitz, H. Teaching the environment to control quantum systems.
{\em Phys. Rev.~A.} {\bf 2006}, {\em 73} (6), 062102, doi: \href{https://doi.org/10.1103/PhysRevA.73.062102}{10.1103/PhysRevA.73.062102}.

\bibitem{HigginsNatComm2014}
Higgins, K.D.B.; Benjamin, S.C.; Stace, T.M.; Milburn, G.J.; 
Lovett, B.W.; Gauger, E.M. Superabsorption of light via quantum engineering. 
{\em Nat. Commun.} {\bf 2014}, {\em 5}, 4705, doi: \href{https://doi.org/10.1038/ncomms5705}{10.1038/ncomms5705}. 

\bibitem{HwangPRA2012}
Hwang, B.; Goan, H.-S. Optimal control for non-Markovian open quantum systems.
{\em Phys. Rev.~A} {\bf 2012}, {\em 85}:3, 032321, doi: \href{https://doi.org/10.1103/PhysRevA.85.032321}{10.1103/PhysRevA.85.032321}.

\bibitem{LucasRPL2013}
Lucas, F.; Hornberger, K. Adaptive Resummation of Markovian Quantum Dynamics.
{\em Phys. Rev. Lett.} {\bf 2013}, {\em 110}:24, 240401, doi: \href{https://doi.org/10.1103/PhysRevLett.110.240401}{10.1103/PhysRevLett.110.240401}.

\bibitem{LiningtonPRA2008}
Linington, I.E.; Garraway, B.M. Dissipation control in cavity 
QED with oscillating mode structures.
{\em Phys. Rev. A} {\bf 2008}, {\em 77}:3, 033831, doi: \href{https://doi.org/10.1103/PhysRevA.77.033831}{10.1103/PhysRevA.77.033831}.

\bibitem{Kozyrev2016} 
Volovich, I.V., Kozyrev, S.V. Manipulation of states of a degenerate quantum system. Proc. Steklov Inst. Math. 294, 241–251 (2016), doi: \href{https://doi.org/10.1134/S008154381606016X}{10.1134/S008154381606016X}.

\bibitem{ZhongPRA2011}
Zhong, H.; Hai, W.; Lu, G.; Li, Z. 
Incoherent control in a non-Hermitian Bose-Hubbard dimer.
{\em Phys. Rev.~A} {\bf 2011}, {\em 84}, 013410, doi: \href{https://doi.org/10.1103/PhysRevA.84.013410}{10.1103/PhysRevA.84.013410}.

\bibitem{SinghPRA2007}
Singh, K.P.; Rost, J.M. Femtosecond photoionization of atoms under noise.
{\em Phys. Rev. A} {\bf 2007}, {\em 76}:6, 063403, doi: \href{https://doi.org/10.1103/PhysRevA.76.063403}{10.1103/PhysRevA.76.063403}.

\bibitem{MukhopadhyayPRA2018}
Mukhopadhyay, C. Generating steady quantum coherence and magic through 
an autonomous thermodynamic machine by utilizing a spin bath.
{\em Phys.~Rev.~A} {\bf 2018}, {\em 98}:1, 012102, doi: \href{https://doi.org/10.1103/PhysRevA.98.012102}{10.1103/PhysRevA.98.012102}.

\bibitem{QutubuddinPRR2021}
Qutubuddin, Md.; Dorfman, K.E. Incoherent control of optical signals: 
Quantum-heat-engine approach. {\em Phys. Rev. Res.}
{\bf 2021}, {\em 3}:2, 023029, doi: \href{https://doi.org/10.1103/PhysRevResearch.3.023029}{10.1103/PhysRevResearch.3.023029}.

\bibitem{LaforgeJCP2018}
Laforge, F.O.; Kirschner, M.S.; Rabitz, H.A. 
Shaped incoherent light for control of kinetics: Optimization of up-conversion hues in phosphors.
{\em J.~Chem. Phys.} {\bf 2018}, {\em 149}, 054201, doi: \href{https://doi.org/10.1063/1.5035077}{10.1063/1.5035077}.

\bibitem{Pechen_Trushechkin_2015}
Pechen, A.N.; Trushechkin, A.S. Measurement-assisted Landau-Zener 
transitions, {\em Phys. Rev. A} {\bf 2015}, {\em 91}:5, 052316, doi: \href{https://doi.org/10.1103/PhysRevA.91.052316}{10.1103/PhysRevA.91.052316}.

\bibitem{Davies1976} 
Davies, E.B. Quantum theory of open systems. {\em Academic Press} (1976).

\bibitem{Accardi2002} 
Accardi, L., Lu;~Y.G., Volovich~I. Quantum Theory and Its Stochastic Limit. {\em Springer} (2002). doi: \href{https://doi.org/10.1007/978-3-662-04929-7}{10.1007/978-3-662-04929-7}.

\bibitem{Trushechkin2021}
Trushechkin, A. Unified Gorini–Kossakowski–Lindblad–Sudarshan quantum master equation beyond the secular approximation, Phys. Rev. A, 103 (2021), 062226, doi: \href{https://doi.org/10.1103/PhysRevA.103.062226}{10.1103/PhysRevA.103.062226}.

\bibitem{Pechen_2011} 
Pechen, A. Engineering arbitrary pure and mixed quantum states.
{\em Phys. Rev.~A} {\bf 2011}, {\em 84} (4), 042106, doi: \href{https://doi.org/10.1103/PhysRevA.84.042106}{10.1103/PhysRevA.84.042106}.

\bibitem{Wu_Pechen_Brif_Rabitz_2007} 
Wu, R.; Pechen, A.; Brif, C.; Rabitz, H. 
Controllability of open quantum systems with Kraus-map dynamics.
{\em J.~Phys.~A: Math.~Theor.} {\bf 2007}, {\em 40}:21, 5681--5693, doi: \href{https://doi.org/10.1088/1751-8113/40/21/015}{10.1088/1751-8113/40/21/015}.

\bibitem{LokutsievskiyJPA2021}
Lokutsievskiy, L.; Pechen, A. Reachable sets for two-level open quantum systems 
driven by coherent and incoherent controls. {\em J.~Phys.~A: Math. Theor.}
{\bf 2021}, {\em 54}, 395304, doi: \href{https://doi.org/10.1088/1751-8121/ac19f8}{10.1088/1751-8121/ac19f8}.

\bibitem{Boscain_Sigalotti_Sugny_2021}
Boscain, U.; Sigalotti, M.; Sugny, D. Introduction to 
the Pontryagin maximum principle for quantum optimal control.
{\em PRX Quantum} {\bf 2021}, {\em 2}:3, 030203, doi: \href{https://doi.org/10.1103/PRXQuantum.2.030203}{10.1103/PRXQuantum.2.030203}.

\bibitem{Gross_Neuhauser_Rabitz_article_1992}
Gross, P.; Neuhauser, D.; Rabitz, H. Optimal control of
curve-crossing systems, {\em J.~Chem. Phys.} {\bf 1992}, {\bf 96} (4),
2834--2845, doi: \href{https://doi.org/10.1063/1.461980}{10.1063/1.461980}.

\bibitem{Tannor_Kazakov_Orlov_1992}
Tannor, D.J.; Kazakov, V.; Orlov, V. Control of photochemical branching:
Novel procedures for finding optimal pulses and global upper bounds.
In {\em Time-Dependent Quantum Molecular Dynamics};
Broeckhove, J., Lathouwers, L., Eds.; Springer: Boston, MA, 1992; pp. 347--360, 
doi: \href{https://doi.org/10.1007/978-1-4899-2326-4_24}{10.1007/978-1-4899-2326-4\_24}. 

\bibitem{Jager_Reich_Goerz_Koch_Hohenester_2014}
J\"{a}ger, G.; Reich, D.M.; Goerz, M.H.; Koch, C.P.; Hohenester, U.
Optimal quantum control of Bose-Einstein condensates in magnetic microtraps: 
Comparison of GRAPE and Krotov optimization schemes.
{\em Phys. Rev.~A} {\bf 2014}, {\em 90}:3, 033628, doi: \href{https://doi.org/10.1103/PhysRevA.90.033628}{10.1103/PhysRevA.90.033628}.

\bibitem{Morzhin_Pechen_RussianMathSurveys_2019} 
Morzhin, O.V.; Pechen, A.N. Krotov method for optimal control of closed
quantum systems. {\em Russian Math. Surveys} {\bf 2019}, {\em 74}, 851--908, doi: \href{https://iopscience.iop.org/article/10.1070/RM9835/meta}{10.1070/RM9835}.

\bibitem{Zhu_Rabitz_1998}
Zhu, W.; Rabitz, H. A rapid monotonically convergent iteration
algorithm for quantum optimal control over the expectation value of
a positive definite operator, {\em J. Chem. Phys.} {\bf 1998}, {\em 109}, 385, doi: \href{https://doi.org/10.1063/1.476575}{10.1063/1.476575}.

\bibitem{Maday_Turinici_2003}
Maday, Y.; Turinici, G. New formulations of monotonically
convergent quantum control algorithms, {\em J. Chem. Phys.} {\bf 2003}, {\em 118}:18, 8191--8196, doi: \href{https://doi.org/10.1063/1.1564043}{10.1063/1.1564043}. 

\bibitem{Khaneja_Reiss_Kehlet_SchulteHerbruggen_Glaser_2005}
Khaneja, N; Reiss, T; Kehlet, C; Schulte-Herbr\"{u}ggen, T; Glaser, S.J.  
Optimal control of coupled spin dynamics: design of NMR pulse sequences 
by gradient ascent algorithms. {\em J.~Magn.~Reson.} {\bf 2005}, {\em 172}:2,
296--305, doi: \href{https://doi.org/10.1016/j.jmr.2004.11.004}{10.1016/j.jmr.2004.11.004}.

\bibitem{SchulteHerbruggen_Sporl_Khaneja_Glaser_2011}
Schulte-Herbr\"{u}ggen, T.; Sp\"{o}rl, A.; Khaneja, N.; Glaser, S.J.  
Optimal control for generating quantum gates in open dissipative systems.
{\em J.~Phys.~B: At. Mol. Opt. Phys.} {\bf 2011}, {\em 44}:15, 154013, doi: \href{https://doi.org/10.1088/0953-4075/44/15/154013}{10.1088/0953-4075/44/15/154013}.

\bibitem{Volkov_Morzhin_Pechen_JPA_2021}
Volkov, B.O.; Morzhin, O.V.; Pechen, A.N. Quantum control landscape for 
ultrafast generation of single-qubit phase shift quantum gates.
{\em J.~Phys.~A: Math. Theor.} {\bf 2021}, {\em 54}, 215303, doi: \href{https://doi.org/10.1088/1751-8121/abf45d}{10.1088/1751-8121/abf45d}.

\bibitem{Judson_Rabitz_1992}
Judson, R.S.; Rabitz, H. Teaching lasers to control
molecules, {\it Phys. Rev. Lett.} {\bf 1992}, {\bf 68} 1500, doi: \href{https://doi.org/10.1103/PhysRevLett.68.1500}{10.1103/PhysRevLett.68.1500}.

\bibitem{Ananevskii_Fradkov_2005}
Anan'evskii, M.S.; Fradkov, A.L. Control of the observables 
in the finite-level quantum systems, {\em Autom. Remote Control}
{\bf 2005}, {\em 66}:5, 734--745, doi: \href{https://doi.org/10.1007/s10513-005-0117-y}{10.1007/s10513-005-0117-y}.

\bibitem{Pechen_RMS_2016}
Pechen A.N. On the speed gradient method for generating unitary quantum operations for closed quantum systems, {\em Russian Math. Surveys}, {\bf 2016}, {\em 71}:3, 597–599, doi: \href{https://iopscience.iop.org/article/10.1070/RM9722}{10.1070/RM9722}.

\bibitem{Caneva_Calarco_Montangero_2011}
Caneva, T.; Calarco, T.; Montangero, S. Chopped random-basis
quantum optimization, {\em Phys. Rev. A} {\bf 2011}, {\bf 84} (2), 022326, doi: \href{https://doi.org/10.1103/PhysRevA.84.022326}{10.1103/PhysRevA.84.022326}.

\bibitem{Dalgaard_Motzoi_et_al_2020}
Dalgaard, M.; Motzoi, F.; Hasseriis, J.; Jensen, M.; Sherson, J.
Hessian-based optimization of constrained quantum control,
{\em Phys. Rev. A} {\bf 2020}, {\em 102}:4, 042612, doi: \href{https://doi.org/10.1103/PhysRevA.102.042612}{10.1103/PhysRevA.102.042612}.

\bibitem{Bondar_Pechen_2020}
Bondar, D.I.; Pechen, A.N. Uncomputability and complexity of quantum control,
{\em Sci. Rep.} {\bf 2020}, {\em 10}:1, 1195, doi: \href{https://doi.org/10.1038/s41598-019-56804-1}{10.1038/s41598-019-56804-1}.

\bibitem{Morzhin_Pechen_IJTP_2021_2019}  
Morzhin, O.V.; Pechen, A.N. Minimal time generation of density 
matrices for a two-level quantum system driven by coherent and 
incoherent controls. {\em Int. J.~Theor. Phys.} {\bf 2021},
{\em 60}, 576--584, doi: \href{https://doi.org/10.1007/s10773-019-04149-w}{10.1007/s10773-019-04149-w}.

\bibitem{Morzhin_Pechen_Physics_of_Particles_and_Nuclei_2020} 
Morzhin, O.V.; Pechen, A.N. Maximization of the Uhlmann--Jozsa 
fidelity for an open two-level quantum system with coherent 
and incoherent controls. {\em Phys. Part. Nucl.} 
{\bf 2020}, {\em 51} (4), 464--469, doi: \href{https://doi.org/10.1134/S1063779620040516}{10.1134/S1063779620040516}. 

\bibitem{Morzhin_Pechen_LJM_2020} 
Morzhin, O.V.; Pechen, A.N. Machine learning for finding suboptimal 
final times and coherent and incoherent controls for an open two-level 
quantum system. {\em Lobachevskii J.~Math.} {\bf 2020}, {\em 41} (12), 
2353--2369, doi: \href{https://doi.org/10.1134/S199508022012029X}{10.1134/S199508022012029X}.

\bibitem{Morzhin_Pechen_SteklovProceedings_2021} 
Morzhin, O.V.; Pechen, A.N. On reachable and controllability sets 
for time-minimal control of an open two-level quantum system.
{\em Proc. Steklov Inst. Math.} {\bf 2021}, {\em 313}, 149--164, doi: \href{https://doi.org/10.1134/S0081543821020152}{10.1134/S0081543821020152}.

\bibitem{Morzhin_Pechen_AIP_Conf_Proc_2021} 
Morzhin, O.V.; Pechen, A.N. Numerical estimation of reachable 
and controllability sets for a two-level open quantum system 
driven by coherent and incoherent controls.
{\em AIP Conf. Proc.} {\bf 2021}, {\em 2362}, 060003, doi: \href{https://doi.org/10.1063/5.0055004}{10.1063/5.0055004}.

\bibitem{Morzhin_Pechen_arXiv2205.02521}  
Morzhin, O.V.; Pechen, A.N. On optimization of coherent and incoherent 
controls for two-level quantum systems. {\em Izvestiya: Mathematics (accepted)}, {\bf 2022}, doi: \href{
https://doi.org/10.48550/arXiv.2205.02521}{10.48550/arXiv.2205.02521}.

\bibitem{Wang_Babikov_2011} 
Wang, L.; Babikov, D. Adiabatic coherent control in the anharmonic 
ion trap: Proposal for the vibrational two-qubit system.
{\em Phys. Rev.~A} {\bf 2011}, {\em 83} (5), 052319, doi: \href{https://doi.org/10.1103/PhysRevA.83.052319}{10.1103/PhysRevA.83.052319}.

\bibitem{Allen_Kosut_Joo_Leek_Ginossar_2017} 
Allen, J.L.; Kosut, R.; Joo, J.; Leek, P.; Ginossar, E.
Optimal control of two qubits via a single cavity drive in circuit quantum 
electrodynamics. {\em Phys. Rev.~A} {\bf 2017}, {\em 95} (4), 042325, doi: \href{https://doi.org/10.1103/PhysRevA.95.042325}{10.1103/PhysRevA.95.042325}.

\bibitem{Hu_Ke_Ji_2018} 
Hu, J.; Ke, Q.; Ji, Y. Steering quantum dynamics of a 
two-qubit system via optimal bang-bang control.
{\em Int.~J. Theor. Phys.} {\bf 2018}, {\em 57} (5), 1486--1497, doi: \href{https://doi.org/10.1007/s10773-018-3676-8}{10.1007/s10773-018-3676-8}.

\bibitem{Bukov_Day_Weinberg_et_al_2018} 
Bukov, M.; Day, A.G.R.; Weinberg, P.; Polkovnikov, A.; Mehta, P.; 
Sels, D. Broken symmetry in a two-qubit quantum control landscape.
{\em Phys. Rev.~A} {\bf 2018}, {\em 97} (5), 052114, doi: \href{https://doi.org/10.1103/PhysRevA.97.052114}{10.1103/PhysRevA.97.052114}.

\bibitem{Morzhin_Pechen_LJM_2021}  
Morzhin, O.V.; Pechen, A.N. Generation of density matrices 
for two qubits using coherent and incoherent controls.  
{\em Lobachevskii J.~Math.} {\bf 2021}, {\em 42}:10, 2401--2412, doi: \href{https://doi.org/10.1134/S1995080221100176}{10.1134/S1995080221100176}.

\bibitem{Holevo_book_2019}
Holevo, A.S. \textit{Quantum Systems, Channels, Information: A Mathematical Introduction}. 
2nd Edition; Berlin, Boston: De Gruyter, 2019, doi: \href{https://doi.org/10.1515/9783110273403}{10.1515/9783110273403}.

\bibitem{Basilewitsch_Koch_Reich_2019}
Basilewitsch, D; Koch, C.P.; Reich, D.M. Quantum optimal 
control for mixed state squeezing in cavity optomechanics.
{\em Adv. Quantum Technol.} {\bf 2019}, {\em 2}, 1800110, doi: \href{https://doi.org/10.1002/qute.201800110}{10.1002/qute.201800110}.

\bibitem{Manko_Manko_Entropy_2021}
Man’ko, O.V.; Man’ko, V.I. Probability representation of quantum states.
{\em Entropy} {\bf 2021}, {\em 23}:5, 549, doi: \href{
https://doi.org/10.1007/s10946-019-09778-4}{10.1007/s10946-019-09778-4}.

\bibitem{Bertlmann_Krammer_2008}
Bertlmann, R.A.;Krammer P. Bloch vectors for qudits. {\em J. Phys. A: Math. Theor.}, {\bf  2008}, {\em 41}, 235303, doi: \href{
https://doi.org/10.1088/1751-8113/41/23/235303
}{10.1088/1751-8113/41/23/235303}.

\bibitem{Wilcox1967}
Wilcox, R.M. Exponential Operators and Parameter Differentiation in Quantum Physics.
{\em J.~Math. Phys.} {\bf 1967}, {\em 8}:4, 962, doi: \href{https://doi.org/10.1063/1.1705306}{10.1063/1.1705306}.

\end{thebibliography}
\end{document}